\PassOptionsToPackage{dvipsnames}{xcolor}
\documentclass[10pt, USenglish, reqno]{article}
\usepackage{amsmath}
\usepackage{amsfonts}
\usepackage{amssymb}
\usepackage{amsthm}
\usepackage[USenglish]{babel}
\usepackage{diagram}
\usepackage{etoolbox}
\usepackage{graphicx}
\usepackage{macro}
\usepackage{mathtools}
\usepackage{nicefrac}
\usepackage{tikz}
\usepackage{xcolor}
\usepackage{hyperref}
\usetikzlibrary{shapes,arrows,positioning,fit}

\hypersetup{%
	colorlinks,
	citecolor=ForestGreen,
	linkcolor=MidnightBlue,
}
{\list{}{\leftmargin=#1\rightmargin=#1}\item[]}%
{\endlist}
\AtEndEnvironment{Example}{\hfill\qedsymbol}{}{}
\AtEndEnvironment{Remark}{\hfill\qedsymbol}{}{}

\tikzset{
	arrow/.style = {
		shorten >= 2pt,
		shorten <= 2pt,
	},
}

\newcommand{\grad}{\nabla}

\title{%
	Nonlinear Sampled-data Systems\\A lifting framework%
	\thanks{%
		This work was supported by JSPS KAKENHI Grant Numbers JP19H02161 and JP20K14766.
		Yutaka Yamamoto also wishes to thank DIGITEO and
		Laboratoire des Signaux et Systemes
		(L2S, UMR CNRS), CNRS-CentraleSupelec-University Paris-Sud and Inria Saclay
		for their financial support while part of this research was conducted.
		This is an author's version of the paper published as Y. Yamamoto and K. Yamamoto, `Nonlinear sampled-data systems—a lifting framework', IFAC-PapersOnLine, vol. 56, no. 2, pp. 6406–6410, 2023.
		Minor typos have been corrected in this version.
	}%
}

\author{%
	Yutaka Yamamoto\thanks{%
		Professor Emeritus,
		Graduate School of Informatics, Kyoto University, Kyoto 606-8501, Japan.
		{\it E-mail:} yy@i.kyoto-u.ac.jp%
	}%
	\and
	Kaoru Yamamoto\thanks{%
		Faculty of Information Science and Electrical Engineering,
		Kyushu University, Fukuoka 819-0395, Japan.
		{\it E-mail:} yamamoto@ees.kyushu-u.ac.jp%
	}%
}

\date{}

\begin{document}

\maketitle

\begin{abstract}
This short note gives a new framework for dealing
with nonlinear sampled-data systems.
We introduce a new idea of lifting, which is well known
for linear systems, but not successfully generalized
to nonlinear systems.
This paper introduces a new lifting technique for
nonlinear, time-invariant systems, which are different
from the linear counterpart as developed in \cite{BPFT91,YYTAC94}
etc.  The main difficulty is that the direct feedthrough term
effective in the linear case cannot be generalized to the nonlinear
case.  Instead, we will further lift the state trajectory,
and obtain an equivalent time-invariant discrete-time system
with function-space input and output spaces.
The basic framework, as well as the closed-loop equation
with a discrete-time controller, is given. As an application
of this framework, we give a representation for the
Koopman operator derived from the given original nonlinear system.
\end{abstract}

\newcommand{\hatp}{\mbox{$\hat{p}$}}
\newcommand{\hatq}{\mbox{$\hat{q}$}}
\newcommand{\hatx}{\mbox{$\hat{x}$}}
\newcommand{\haty}{\mbox{$\hat{y}$}}
\newcommand{\hatphi}{\mbox{$\hat{\phi}$}}
\newcommand{\hatpsi}{\mbox{$\hat{\psi}$}}
\newcommand{\hatg}{\mbox{$\hat{g}$}}
\newcommand{\hatG}{\mbox{$\hat{G}$}}

\newcommand{\lift}{\mbox{$\mathcal{L}$}}
\newcommand{\lifta}[1]{\mbox{$\mathcal{L}[#1]$}}
\newcommand{\de}{{\rm d}}
\NewDocumentCommand{\deriv}{s o m}{%
	\IfBooleanTF{#1}{\tfrac}{\frac}
	{\de\IfValueT{#2}{^{#2}}#3}{\de t\IfValueT{#2}{^{#2}}}
}


\section{Introduction}
The lifting technique \cite{BPFT91,YYTAC94} has played
a crucial role in modernizing the treatment of
linear sampled-data control systems, e.g.,
\cite{YYencyclopedia,YYencyclopedia2}.
To see the basic idea, let us take the following
linear, time-invariant system:
\begin{equation}
\begin{aligned} \label{eqn:linsys}
\deriv*{}x(t) & = Ax(t) + Bu(t) \\
			y(t) & = Cx(t)
\end{aligned}
\end{equation}
where $x, u, y$ are the state, input, and output, respectively,
and $A, B, C$ are constant matrices of suitable dimensions.

The {\em lifting\/} converts this system to a linear,
time-invariant, {\em discrete-time\/} system as follows.
Fix an arbitrary $T>0$ (which may later work as a
sampling period), and let $\lift$ be the {\em lifting\/}
operator defined by
\begin{equation*}
\lift: \psi \mapsto \{\psi_{k}(\cdot) \},
	\psi_{k}(\theta) := \psi(kT+\theta),
\end{equation*}
for $0\leq \theta < T$ and $\psi$ belonging to a suitably defined function space
on $[0, \infty)$, e.g., $\Ltwo$ or $\Ltwoloc$, etc.
The lifting employed in \cite{BPFT91} gives
\begin{equation}
\begin{aligned} \label{eqn:lifted1}
x[k+1] & =
e^{AT}x[k] + \int_{0}^{T} e^{A(T-\tau)}Bu[k](\tau)d\tau, \\
y[k](\theta) & =
Ce^{A\theta}x[k] + \int_{0}^{\theta} Ce^{A(\theta-\tau)}Bu[k](\tau)d\tau.
\end{aligned}
\end{equation}
where $x[k]$ denote the state $x(t)$ at time $kT$.
These formulas now take the form
\begin{equation}
\begin{aligned} \label{eqn:linearlifting}
x[k+1] & = {\mathcal A} x[k] + {\mathcal B} u[k] \\
	y[k] & = {\mathcal C} x[k] + {\mathcal D} u[k].
\end{aligned}
\end{equation}

Observe that the operators
${\mathcal A}, {\mathcal B}, {\mathcal C}, {\mathcal D}$
do not depend on the time variable $k$, and hence the above
is a time-invariant discrete-time system, where $u[k]$ and
$y[k]$ belong to a suitable function space, say, $\Ltwoot$.

A crucial point here is the direct feedthrough term
$\mathcal D$, which does not exist in the original
continuous-time system.  This term describes the
effect of the input to the output, which, in principle,
should depend on the state evolution in the mean time.
However, the linearity of the original system makes it
possible to describe this effect without involving
the state transition.  When we deal with a nonlinear
system, this is clearly impossible, and this is the
topic of the present article.

\section{Nonlinear system lifting}

Suppose we are given the following continuous-time nonlinear
system:
\begin{equation}
\begin{aligned} \label{eqn:nonlinearsystem}
	\deriv*{}x(t) & = f(x(t), u(t)), \\
				y(t) & = h(x(t)), \\
				x(0) & = x_{0},
\end{aligned}
\end{equation}
where $x(t) \in \R^{n}$ and we assume a suitable
condition for $f$ to allow for the existence and
uniqueness for the solution to exist, for example,
the Lipschitz condition with a global
constant not to allow a finite escape time.
We also assume a suitable regularity condition for $h$.  The
initial condition is $x(0) = x_{0}$.

Let $\Phi(t, x_{0}, u)$ denote the solution
$x(t)$ at time $t$ of the above differential
equation.
Let $T > 0$ be fixed; this will be taken as a
sampling period later.
We then define the following nonlinear discrete-time
system:
\begin{equation}
\begin{aligned} \label{eqn:liftedsys}
	x[k+1](\theta) & = \Phi\left((k+1)T+\theta, x[k](T), u[k+1](\cdot)\right) \\
	y[k](\theta) & = h(x[k](\theta)).
\end{aligned}
\end{equation}
Here, we mean by $u[k+1](\cdot)$ that $x[k+1](\theta)$
depends on $u[k+1](\tau),$ $0\leq \tau \leq \theta$.
We call this system the {\em lifting\/} of the system
\eqref{eqn:nonlinearsystem}.  Note that it
does not satisfy the strict causality condition
in that $u[k+1]$ appears on the right-hand side.

\begin{Remark}
To be precise, \eqref{eqn:liftedsys} requires a little
more care.  When lifted, $x[k](\theta)$ is defined
only for $0 \leq \theta < T$, and hence not for $T$.
To circumvent this, we should require that the
trajectory $x(t)$ of \eqref{eqn:nonlinearsystem}
is continuous, and interpret that $x[k](T)$
as $\lim_{\theta \uparrow T} x[k](\theta)$.
Also, the point evaluation $h(x[k](\theta))$ at
$t = kT+\theta$ is not necessarily well defined.
To take care of this, we should assume that
$h$ is defined on a dense ${\cal D}(h)$ subspace of the
state space, and the mapping
\begin{equation}
h: {\cal D}(h) \rightarrow \C
\end{equation}
gives rise to a continuous mapping
\begin{equation}
\tilde{h}: {\cal D}(h) \rightarrow \Ltwoloc: x \mapsto h(x(t)).
\end{equation}
We will write $h(x(t))$ as a shorthand for
$\tilde{h}(x)(t)$.  By the assumed continuity, this
mapping $\tilde{h}$ can be extended continuously to the
whole state space $\R^{n}$.  For details on this extension,
see \cite{YYReal81}.
\end{Remark}

To see the difference from the usual linear case,
consider the following example.
\begin{Example}
Consider the linear system \eqref{eqn:linsys}.
Then \eqref{eqn:linsys} takes the following form:
\begin{align*}
x[k+1](\theta) & = e^{A\theta}x[k](T) +
				\int_{0}^{\theta}e^{A(\theta-\tau)}u[k+1](\tau)d\tau \\
y[k](\theta) & = Cx[k](\theta). \nonumber
\end{align*}
Note that this looks very different from \eqref{eqn:lifted1}
in that the state $x(t)$ is also lifted.
This is the lifting employed in \cite{YYTAC94}.  In this
linear case, one can make it strictly causal by introducing
the new state variable $x[k] - {\mathcal B}u[k]$.  But
this technique cannot be applied to the nonlinear case
considered here.
\end{Example}

\section{Fast-sampling approximation}

The difference of the new nonlinear lifting \eqref{eqn:liftedsys}
from the linear one is that we cannot have a closed-form expression as
\eqref{eqn:linearlifting}.  This is inevitable at the abstract level,
but one can instead introduce a fast-sampling approximation formula as
in the case with the linear case.

The idea is similar to the linear case.  We introduce the subdivision
of the sampling interval $[0, T]$ at each sampling period as
$[0, T/N), [T/N, 2T/N), \ldots, [T-T/N, T)$.
In view of the differential equation \eqref{eqn:nonlinearsystem},
we have
\begin{equation*}
x[0](T/N) - x_{0}
	= \int_{0}^{\nicefrac{T}{N}} f(x(\tau),u[0](\tau))d\tau.
\end{equation*}
One can invoke a suitable approximation for this
equation, e.g., the Runge-Kutta method, or even forward-difference
approximation.  Repeating this process inductively, we
obtain an approximant
of $x[k](T/N), x[k](2T/N), \ldots, X[k](T)$ in \eqref{eqn:liftedsys},
which gives an approximation of
$\Phi(\cdot, x_{0}, u)$ in \eqref{eqn:liftedsys}.  It can
be used as a substitute of the lifted nonlinear system
\eqref{eqn:liftedsys}.  Note here that we do not require
the actual sampling of the trajectory, but it is only
an artificial approximation of the lifted system.

\section{Closed-loop equation}
\label{sec:clloop}
\begin{figure}[tb]
\centering
\tikzset{
        block/.style = {draw, rectangle, thick,
            minimum height=1cm,
            minimum width=1cm},
        input/.style = {coordinate,node distance=1cm},
        output/.style = {coordinate,node distance=1cm},
        arrow/.style={draw, -latex,node distance=2cm},
        pinstyle/.style = {pin edge={latex-, black,node distance=2cm}},
        sum/.style = {draw, circle, node distance=1cm},
    }%
\begin{tikzpicture}[auto, node distance=1.5cm,>=latex]
    \node[input] (input) {};
    \node[sum, right of = input] (sum) {};
    \node[block, right of = sum] (controller) {$C$};
    \node[block, right of = controller] (hold) {$\mathcal H$};
    \node[block, right of = hold] (system) {$P$};
   \node [output, right = 1.5cm of system] (output) {};
    \draw [->] (controller) -- node[name=v]{$v$}(hold);
    \draw [->] (hold) -- node[name=u]{$u$}(system);
    \draw [draw,->] (input) -- node[xshift=-3mm]{$r$} (sum);
    \draw [-] (sum)-- ++(0.3,0) -- node {$e$} ++ (0.4,0.2);
    \draw [->, shorten >=1pt] (sum)++(0.7,0) -- (controller);
    \draw [->, shorten >=1pt] (system) -- node [name=y] {$y$}(output);
    \draw [->, shorten >=1pt] (y) -- ++ (0,-2) -| node [pos=0.96] {$-$} (sum);
    \node at (sum)[xshift=-2.5mm,yshift=2mm] {$+$};
    \fill (y.south) circle [radius=1.5pt];
\end{tikzpicture}
\label{Fig:unifeedback}
\caption{Unity feedback nonlinear sampled-data system}
\end{figure}
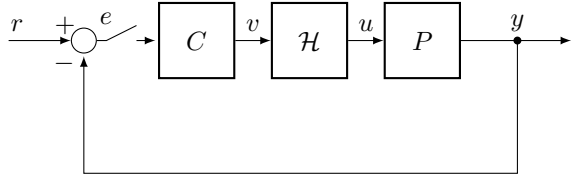

Consider the unity feedback sampled-data control system
in Fig.~\ref{Fig:unifeedback}.
The plant $P$ is the nonlinear continuous-time system
\begin{equation}
\begin{aligned} \label{eqn:plant}
\deriv*{}x(t)& = f(x, u) \\
	y(t) & = h(x).
\end{aligned}
\end{equation}
The controller $C$ is a discrete-time system
\begin{equation*}
\begin{aligned} \label{eqn:controller}
	z[k+1] & = \phi(z[k], e[k](0)) \\
	u[k] & = \psi(z[k]),
\end{aligned}
\end{equation*}
where $z[k], e[k], v[k]$ belong to $\R^{n_{c}}, \R^{m_{c}}, \R^{p_{c}}$,
respectively.

As before, \eqref{eqn:plant} is lifted as \eqref{eqn:liftedsys}:
\begin{align*}
x[k+1](\theta) & = \Phi((k+1)T+\theta, x[k](T), u[k+1](\cdot)) \\
	y[k](\theta) & = h(x[k](\theta)).
\end{align*}
Equating $u[k+1](\cdot) = v[k]$ considering the delay
induced by a strictly proper controller $C$, we obtain the
closed-loop equation
\begin{equation}
\begin{aligned} \label{eqn:closedloop}
	x[k+1](\theta) & = \Phi((k+1)T+\theta, x[k](T), v[k]) \\
	z[k+1] & = \phi(z[k], e[k](0)) \\
	v[k] & = \psi(z[k]) \\
	e[k](\theta) & = r[k](\theta) - y[k](\theta).
\end{aligned}
\end{equation}
In the above equation, the lack of strict causality disappears.

\section{The Koopman operator} \label{sec:Koopman}

The Koopman operator \cite{Koopman1931} has been actively studied for
nonlinear systems; see, for example, \cite{MauroyMezicSusuki2020}.
The crux of the idea is in introducing
the time-transition of the {\em observables}.
That is, considering the adjoint (dual) system of the
original system.
While the original system may be nonlinear, the
transition of the adjoint system turns out to be
linear, and this is a great advantage of the Koopman
operator.

\subsection{Some backgrounds}

While the study of the Koopman operator has
become popular relatively recently, the very fact
that the introduction of a duality for a nonlinear
object has been long known.  Let us note some of
this background.

R. E. Kalman had discussed a lot about dual systems,
particularly duality between reachability (controllability)
and observability \cite{CIME1968,KFA69}, and this idea was further explored
in the context of realization theory; see, e.g.,
\cite{Sontagthesis, SontagMathControlTheory,YYReal81}.
The system derived from the
evolution of observables is called the dual (adjoint) system, and
this is precisely the Koopman operator idea.
Various observability notions can be translated into
some system properties (e.g., reachability) by going into
duals.  For example, in \cite{Sontagthesis}, a
class of systems in an algebraic category is studied;
a system is {\em algebraically observable\/} if the
algebra generated by the observables agrees with the
algebra of polynomials.  It is algebraically observable
if and only if its adjoint system satisfies a strong
reachability condition.
In the context of continuous-time linear
systems, a system is {\em topologically observable\/}
if and only if its dual system is exactly reachable
\cite{YYReal81}.
These recognitions date back to the intuition shown in
R. E. Kalman's idea given in \cite{CIME1968}
where a finite-dimensional linear system is
reachable if and only if its dual (or adjoint) system
is observable, and vice versa.

\subsection{The Koopman operator for nonlinear systems}

We now proceed to consider
the system \eqref{eqn:nonlinearsystem}
without input $u$ for simplicity:
\begin{equation}
\begin{aligned} \label{eqn:nonlinearwithzeroinput}
\deriv*{}x(t) & = f(x(t)), \\
			y(t) & = h(x(t)), \\
			x(0) & = x_{0}.
\end{aligned}
\end{equation}
We will see the correspondence
of spectra between the original system and its lifted counterpart.
Under this assumption, we denote $\Phi(t, x, 0)$ by
$\Phi(t)(x)$.  $\Phi(t)$ satisfies the semigroup property:
$\Phi(t+s) = \Phi(t)\Phi(s)$, $\Phi(0) = I$, where the
product is understood as the composition.

The Koopman operator for \eqref{eqn:nonlinearwithzeroinput} is
defined by
\begin{equation} \label{eqn:Koopman1}
(U(t)h)(x) := h(\Phi(t)x).
\end{equation}
We may also write $\innprod{U(t)h}{x}$ for $(U(t)h)(x)$
in view of the linearity of $U(t)$ \cite{MauroyMezicSusuki2020}.
Then \eqref{eqn:Koopman1} can be written as
$\innprod{U(t)h}{x} = h(\Phi(t)x)$.

The lifted version of \eqref{eqn:nonlinearwithzeroinput}
is given by
\begin{equation}
\begin{aligned}\label{eqn:simplelifted}
x[k+1](\cdot) & = \Phi(T,x[k]) = \Phi(T)x[k] \\
	y[k] & = h(x[k]).
\end{aligned}
\end{equation}
Then the discrete-time Koopman operator $U_{d}$ corresponding to
the lifted system \eqref{eqn:simplelifted}
is defined as
\begin{equation} \label{eqn:Koopman2}
U_{d}(h)(x) := h(\Phi(T)(x)).
\end{equation}

It is clear that $U_{d}(h) = U(T)(h)$.

\begin{Example} \rm
\label{ex:linear}
Let us consider the simple case of a continuous-time
linear system.
Consider the linear differential equation
given in a Banach space $X$:
\begin{equation} \label{eqn:linevolution}
\deriv*{}x(t) = Ax(t), x \in X
\end{equation}
and an observation operator
\begin{equation*} \label{eqn:linobservable}
H: X \rightarrow \C: x \mapsto H(x).
\end{equation*}
We here assume the following:
\begin{enumerate}
\item $X$ is a reflexive Banach space;
\item $A$ gives rise to a strongly continuous semigroup
$S(t)$; that is, $S(t) = e^{At}$.
\item For simplicity, we assume $H$
to be a continuous linear functional:
$X \rightarrow \C$.
\end{enumerate}
Since $X$ is reflexive, there exists a dual semigroup
$S'(t)$ which satisfies
\begin{equation*}
\innerprod{S'(t)H}{x} = \innerprod{H}{S(t)x}
\end{equation*}
For every $H \in X', x \in X$.  This is nothing
but the definition of the Koopman operator \eqref{eqn:Koopman1},
and hence $S'(t) = U(t)$.  If we denote the
infinitesimal generator (see Subsection \ref{subsec:spec} below)
of $S(t)$ by $A$, then
the generator of $U(t)$ is given by $A'$ or $A^{*}$,
the adjoint operator of $A$.
\end{Example}
Naturally, in this case, the lifted Koopman operator $U_{d}$
agrees with $S'(T)$.

\begin{Remark} \rm
It is possible to weaken the third requirement
on $H$ above that it is defined on a dense subspace $D(H) \subset X$,
and the mapping
\[
D(H) \rightarrow {\cal F}: x \mapsto HS(t)x
\]
is extensible to a continuous linear map from $X$ to $\cal F$,
where $\cal F$ is a suitably defined function space on
$[0, \infty)$.  In this case, the definition of the
Koopman operator needs to be modified a little,
but the formality works mutatis mutandis. See, e.g.,
\cite{YYReal81} for such an extension.
\end{Remark}

\subsection{Spectrum of the Koopman operator}
\label{subsec:spec}
Let us assume that $U(t)$ is strongly continuous.
The infinitesimal generator of the semigroup $U(t)$ is
then defined by
\begin{equation*} \label{eqn:infgen}
L := \lim_{t\rightarrow 0} \frac{U(t) - I}{t}.
\end{equation*}
It is important to realize that $U(t)$, and hence $L$ also
is a linear operator despite the fact that $f$ is nonlinear.

Let us calculate $Lh$, where $h$ is an observable.
By the chain rule of differentiation, we have
\begin{equation*} \label{eqn:infinitesimal}
\innprod{\frac{1}{t}(U(t)h - h)}{x} =
	(\frac{1}{t}(h(\Phi(t)x) - h(x))
	\rightarrow \grad h\dot{x} = \grad h f(x).
\end{equation*}
This should hold for any $x$, and hence
$Lh = \grad h f$.
(We will see below a more explicit
expression when the system is linear.)

On the other hand, since $L$ is the infinitesimal generator of the strongly
continuous semigroup $U(t)$, we may write $U(t) = e^{Lt}$ for
$t \geq 0$.  Since $U(h)$ is equal to $(U(t)h)_{t=T}$
as noted above, we have that $U_{d}(h) = e^{LT}h$.

Now let us compare the eigenvalues of the
Koopman operator \eqref{eqn:Koopman1} and its lifted
version \eqref{eqn:Koopman2}.  We have the following
proposition:
\begin{Prop}
Let $\lambda$ be an eigenvalue of $L$ with
a corresponding eigenfunction $\phi$:
\begin{equation*}
L \phi = \lambda \phi.
\end{equation*}
Then
\begin{equation} \label{eqn:Uteigen}
U(t) \phi = e^{\lambda t} \phi.
\end{equation}
In particular, the Koopman operator $U_{d}$ of the
lifted system \eqref{eqn:liftedsys} has an eigenvalue
$e^{\lambda T}$.
\end{Prop}
\Proof
Since $U(t) = e^{Lt}$, \eqref{eqn:Uteigen} follows
from the spectral mapping theorem \cite[Chapter IX]{Yosida78}.
The second claim is obvious because $U_{d} = U(T)$.
\EndProof

\begin{Example} \rm
Let us return to the linear case Example \ref{ex:linear}.
Let us assume that $\dim X = n < \infty$ for simplicity.
Then the state transition derived by \eqref{eqn:linevolution}
is simply
\[
x(t) = e^{At}x
\]
for the initial state $x$.  Then the Koopman operator
$U(t)$ must be its dual, and hence is given by
\begin{equation*} \label{eqn:linKoopman}
U(t) = e^{A^{\top}t},
\end{equation*}
where $A^\top$ denotes the transpose of $A$.
If $\lambda$ is an eigenvalue of $A$, then it is also
an eigenvalue of $A^{\top}$, and hence
\[
U(t)\phi = e^{\lambda t}\phi
\]
for some $\phi$ as expected.  In particular, this implies
$e^{\lambda T}$ is an eigenvalue of the lifted
system \eqref{eqn:liftedsys}.

This correspondence can be viewed from a more general
viewpoint.
\end{Example}

Let us view the above correspondence from a different
angle.  The infinitesimal generator $L$ of $U(t)$ is
given by $L = \grad h f$.  In the present case,
$\grad h = h$, $f = A$.  Noting
\[
\innprod{Lh}{x} = \innprod{h}{Ax} = \innprod{A^{\top}h}{x},
\]
for every $x \in X$, we have $L = A^{\top}$ again, as expected.

\subsection{Eigenfunctions}

Eivenfunctions play an important roles in the Koopman mode expansion.
Take the linear equation \eqref{eqn:linevolution}, and let
$\lambda$ be an eigenvalue of $A$ with associated eigenfunction
$\phi$.  The associated Koopman operator is $e^{A^{\top}t}$,
along with the same eigenvalue and left eigenfunction
$\phi^{\top}$.  That is,
\[
\phi^{\top}A^{\top} = \lambda\phi^{\top}.
\]
This readily yields
\begin{equation*}
\phi^{\top}U(t) = \phi^{\top}e^{A^{\top}t} = e^{\lambda t}\phi^{\top}.
\end{equation*}
In particular, for the lifted system,
\begin{equation*}
\phi^{\top}U_{d} = \phi^{\top}U(T) = \phi^{\top}e^{A^{\top}T} = e^{\lambda T}\phi^{\top}.
\end{equation*}

\section{Concluding remarks}

We have introduced a lifting framework for time-invariant
nonlinear systems.  This can have impacts on the study of
nonlinear sampled-data systems, especially when one needs
to discuss intersampling behavior.  For some pertinent
studies, see, e.g., \cite{NesicTeel2007} and references therein.
While the introduced lifted system \eqref{eqn:liftedsys}
is not strictly causal as a discrete-time system, this
drawback can be surpassed by considering its connection
with a discrete-time controller \eqref{eqn:closedloop}
as seen in Section \ref{sec:clloop}.

We have also given a straightforward calculation for the
Koopman operators for the system and the corresponding
lifted system, where the input term is absent.  The
generalization to the case with the input term is
attempted by viewing $u$ as part of the state:
see, e.g.,  \cite{MauroyMezicSusuki2020,BruntonKutz2019}.

\section*{Acknowledgement}
The authors would like to thank Eduardo Sontag for
drawing their attention to the Koopman operator and
its relationship with adjoint systems, and also
for several constructive comments.

\bibliographystyle{plain}
\bibliography{yamamotorefs}

\begin{thebibliography}{10}

\bibitem{BPFT91}
Bassam Bamieh, J~Boyd Pearson, Bruce~A Francis, and Allen Tannenbaum.
\newblock A lifting technique for linear periodic systems with applications to
  sampled-data control.
\newblock {\em Syst. Control Lett.}, 17(2):79--88, 1991.

\bibitem{BruntonKutz2019}
Steven~L Brunton and J~Nathan Kutz.
\newblock {\em Data Driven Science and Engineering---Machine Learning,
  Dynamical Systems, and Control}.
\newblock {Cambridge University Press}, Cambridge, UK, 2019.

\bibitem{CIME1968}
Rudolf~E Kalman.
\newblock {\em Lecture on Controllability and Observability}.
\newblock CIME Summer Course, Roma, 1968.

\bibitem{KFA69}
Rudolf~E Kalman, Peter~L Falb, and Michael~A Arbib.
\newblock {\em Topics in Mathematical System Theory}.
\newblock McGraw-Hill, New York, 1969.

\bibitem{Koopman1931}
B~O Koopman.
\newblock Hamiltonian systems and transformations in hilbert space.
\newblock {\em {SIAM} J. Aplied Dynamical Systems}, 17--1:315--318, 1931.

\bibitem{MauroyMezicSusuki2020}
Alexandre Mauroy, Igor Mezi\'{c}, and Yoshihiko Susuki.
\newblock {\em The Koopman Operator in Systems and Control}.
\newblock {Springer}, New York, 2020.

\bibitem{NesicTeel2007}
Dragan Nesi\'{c} and Andrew~R Teel.
\newblock Sampled-data control of nonlinear systems: an overview of recent
  results.
\newblock In {\em Perspectives in Robust Control; Reza Moheimani Ed., Lecture
  Notes in Control and Information}, volume 268, pages 221--239, {Berlin},
  2007. Springer-Verlag.

\bibitem{Sontagthesis}
Eduardo~Daniel Sontag.
\newblock {\em Polynomial Response Maps}.
\newblock {Springer-Verlag}, Berlin, 1979.

\bibitem{SontagMathControlTheory}
Eduardo~Daniel Sontag.
\newblock {\em Mathematical Control Theory}.
\newblock {Springer-Verlag}, Berlin, 1998.

\bibitem{YYReal81}
Yutaka Yamamoto.
\newblock Realization theory of infinite-dimensional linear systems, parts {I}
  and {II}.
\newblock {\em Math.~Syst.~Theory}, 15:55--77 and 169--190, 1981.

\bibitem{YYTAC94}
Yutaka Yamamoto.
\newblock A function space approach to sampled-data control systems and
  tracking problems.
\newblock {\em IEEE Trans. Autom. Control}, AC-39(4):703--713, 1994.

\bibitem{YYencyclopedia}
Yutaka Yamamoto.
\newblock Digital control.
\newblock In {\em Wiley Encyclopedia of Electrical and Electronics
  Engineering}, volume~5, pages 445--457. John Wiley, 1999.

\bibitem{YYencyclopedia2}
Yutaka Yamamoto and Masaaki Nagahara.
\newblock Digital control.
\newblock In {\em Wiley Encyclopedia of Electrical and Electronics Engineering,
  J. Webster Ed.}, pages 1--19. John Wiley, 2018.

\bibitem{Yosida78}
K{\^o}saku Yosida.
\newblock {\em Functional Analysis}.
\newblock Springer, New York, fifth edition, 1978.

\end{thebibliography}

\end{document}